# DEEP LEARNING SENTIMENT ANALYSIS OF AMAZON.COM REVIEWS AND RATINGS

Nishit Shrestha and Fatma Nasoz

Department of Computer Science, University of Nevada Las Vegas, Las Vegas, Nevada, USA

## ABSTRACT

*Our study employs sentiment analysis to evaluate the compatibility of Amazon.com reviews with their corresponding ratings. Sentiment analysis is the task of identifying and classifying the sentiment expressed in a piece of text as being positive or negative. On e-commerce websites such as Amazon.com, consumers can submit their reviews along with a specific polarity rating. In some instances, there is a mismatch between the review and the rating. To identify the reviews with mismatched ratings we performed sentiment analysis using deep learning on Amazon.com product review data. Product reviews were converted to vectors using paragraph vector, which then was used to train a recurrent neural network with gated recurrent unit. Our model incorporated both semantic relationship of review text and product information. We also developed a web service application that predicts the rating score for a submitted review using the trained model and if there is a mismatch between predicted rating score and submitted rating score, it provides feedback to the reviewer.*

## 1. INTRODUCTION

Sentiment analysis is the task of computationally identifying and categorizing the sentiment expressed by an author in a piece of text. It has a wide range of applications in industry from forecasting market movements based on sentiment expressed in news and blogs, to identifying customer satisfaction and dissatisfaction from their reviews and social media posts. It also forms the basis for other applications like recommender systems.

Today, most e-commerce websites have a separate section where their customers can post reviews for products or service. Important information like customers' opinion on products, reasons for negative reviews, suggestions, etc., can be extracted from the posted reviews by performing sentiment analysis on them. Consumers can also assign a numerical value (i.e., rating) to the product or service they are reviewing. On Amazon.com the rating can be between 1 and 5 where 1 is the worst and 5 is the best. In some instances, there is a mismatch between a customer's review and rating. It is important to identify the reviews with mismatched ratings since individual ratings are used to compute the average rating.

Conventionally, a review text was converted to fixed-length feature vector using bag-of-words or bag-of-n-grams and these feature vectors were later used to train a shallow classifier such as naïve Bayes or support vector machine [1]. Although bag-of-words performed surprisingly well and was popular for many years, it has two major flaws. It loses ordering of words and doesn't consider the semantic relationship between words. For example, words such as "bad", "worst", and "Las Vegas" are equally distant despite the fact that "bad" should be semantically closer to "worst" than "Las

Vegas". Although bag-of-n-grams somewhat considers word order in short context, it suffers from data sparsity and high dimensionality [2].

Recently, deep learning has shown promising results in the field of sentiment analysis. They have been shown to outperform bag-of-words approach when it comes to feature generation. Mikolov et al. [3] introduced an efficient way of representing words in vector space by predicting the current word given context and by predicting surrounding words given current word. Their proposed unsupervised algorithm was able to embed words in a continuous vector space where semantically similar words are mapped to nearby points.

Mikolov et al. also [4] introduced word vectors, which is an unsupervised algorithm to efficiently capture semantics of words. Word vectors represent words as vectors and semantically similar words are closer to each other in vector space. In other words, similar words such as "good" and "great" are closer to each other and are far apart from words like "quick", etc. Interestingly, this vector representation of text also captures linguistic patterns. For example, the result of vector calculation vec("King") - vec("Queen") + vec("Woman") will output a vector that is closer to the vector representation of word "Man" than to any other word. Their work was made available as an open-source project titled word2vec and can be found on Google Code [5].

Socher et al. [6] proposed a recursive neural tensor network for semantic compositionality over a sentiment tree bank. Their proposed model achieved state-of-the-art result for binary sentiment classification of Stanford sentiment tree bank.

Kim [7] designed a simple yet powerful convolutional neural network (CNN) that achieved a very good performance across different datasets for the task of sentiment analysis. The input layer of this network comprised of concatenated word2vec word embeddings, which is then followed by a convolutional layer with multiple filters, then a max pooling layer and finally a softmax layer. Johnson et al. [8] used a variation of bag-of-words model to create feature vectors instead of using low-dimensional word vectors as input to the convolutional neural network (CNN). They also experimented with parallel CNN that had two or more convolutional layers in parallel to learn multiple types of embeddings from one hot input vector.

Ouyang et al. [9] used a seven-layer convolutional neural network architecture to perform sentiment analysis on movie reviews collected from rottentomatoes.com. Their architecture consisted of three convolutional layers, three max-pooling layers and a fully-connected layer with softmax activation.

Following on previous work of Mikolov et al. [3][4][5], Le et al. [2] developed paragraph vectors, which is also an unsupervised algorithm and learns fixed-length feature vectors for variable length text such as sentences, paragraphs, or an entire document. Paragraph vectors has been shown to outperform bag-of-words models and other form of text representation and has achieved state-of-the-art results in several classification tasks including sentiment analysis. Since they are faster to train and have high efficiency in capturing syntactic and semantic relationship of a text, our model used paragraph vectors to learn low-dimensional vector of review text.

In the past, most researchers were interested in only the syntactic and semantic relationship between review texts. Currently, there is a growing interest in adding user and product information to strengthen the feature vector. Tang et al. [10] proposed a neural network model that not only captures the semantics of the review text but also user information that expresses that sentiment. A user was represented as a continuous matrix and the product of word matrix and user matrix was used for sentiment classification using a neural network.

Tang et al. [11] further used a neural network model to embed both user and product information into a vector representation of a document. They first converted a document text to continuous vector representation using a convolutional neural network (CNN). Each user and product were represented as a continuous matrix and the concatenation of user-text and product-text was fed to a CNN for sentiment analysis. IMDB, Yelp 2014, and Yelp 2013 datasets were used for sentiment analysis using their model. They were able to achieve state-of-the-art results with all three datasets.

Chen et al. [12] used a similar one-layer convolutional neural network (CNN) to learn review embeddings of IMDB and Yelp datasets. Their convolutional neural network took reviews of varying length and produced 300-dimensional vectors. To handle varying lengths of reviews, they padded shorter reviews with zero vectors. Filters of width 3 and 5 moved on the word embeddings to perform one-dimensional convolution and produced multiple feature maps. Only useful features were captured by using max-overtime pooling in pooling layer. The output of multiple filters was then concatenated to create a 300-dimensional vector. Softmax function was used as an activation function to train the network over K-classes. They then used a Recurrent Neural Network (RNN) to learn the temporal information as well as to capture both product and user information and reported state-of-the-art results on IMDB and Yelp datasets. Their work is based on the argument that a product that receives positive reviews initially is likely to get more positive reviews in the future and vice-versa.

Motivated by the results of Chen et al.'s model [13], along with syntactic and semantic relationships, our model also learned product and temporal relations of reviews. We believe that these are important features and can significantly improve sentiment classification accuracy. Our work is based on the argument that a popular product might receive more number of higher ratings and a less popular product might receive higher number of lower ratings. We used recurrent neural networks (RNN) to capture such information as they are great in representing sequences.

## 2. METHODOLOGY

In order to analyse the sentiment of Amazon.com reviews we built a model using recurrent neural networks (RNN) with gated recurrent unit (GRU) that learned low-dimensional vector representation of reviews using paragraph vectors and product embeddings.

We first converted Amazon.com product reviews to fixed-length feature vectors using paragraph vectors. These feature vectors were then grouped by product and sorted in temporal order. Each group was used to train an RNN with GRU. The vectors generated in the penultimate layer of the RNN are called product embeddings. These embeddings capture important information like product qualities and temporal relations among reviews. We then concatenated product embeddings with fixed-length vectors generated by paragraph vectors and trained a support vector machine. We also developed a user interface to tackle the review-rating mismatches. There are situations where a user may write a highly positive review but give it 1 or 2 stars or write a highly negative review but give it 4 or 5 stars [13]. Although such cases are rare, they create confusion among users who read these reviews. To address this issue, we labeled reviews with rating 1 and 2 are as 'negative', reviews with rating 3 as 'neutral', and reviews with rating 4 and 5 as 'positive'. We then developed a web service application that uses our classifier to predict a class from the user review. If the predicted class and the class that the submitted rating belongs to are different a warning is submitted to the user so that they can review and correct their rating. Figure 1 illustrates our overall approach.

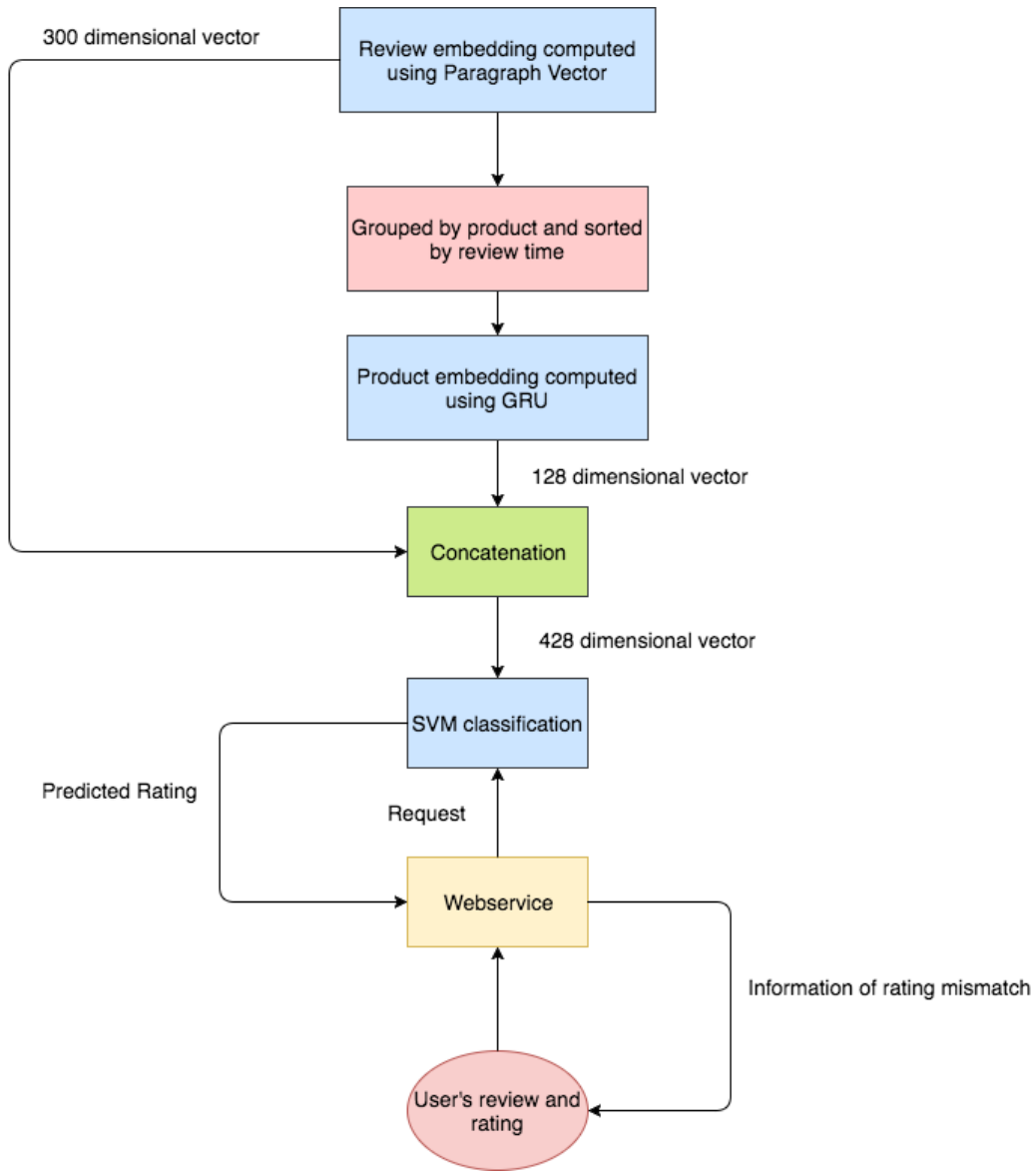

Figure 1 Sentiment analysis of Amazon.com reviews and ratings

## 2.1. DATA AND DATA PRE-PROCESSING

The data used in this study is a set of approximately 3.5 million product reviews collected from Amazon.com by Fang et al. [14]. Each review includes information on rating, product id, helpfulness, reviewer id, review title, review time, and review text. The rating is based on a 5-star scale. Figure 2 displays a review in the dataset.

```
rating: 5.0 out of 5 stars
product_ID: B00DS842HS
helpfulness: 4/4
ID: A28R8UNBXGLFOR
review_by: Melliemel
title: It's working!
review_time: 20140308
review: So far so good. I bought this because I wanted to start oil pulling. It's
    been working great. Great taste (while swishing it around and NOT
    swallowing it). Put some on my arm that was very dry. It helped. Haven't
    cooked with it yet, but I'm sure it will be great!
```

Figure 2 Example Amazon.com review

As shown in Figure 3, there are 2.2 million 5-star reviews; 622,308 4-star reviews; 265,684 3-star reviews; 171,153 2-star reviews; and 288,789 1-star reviews.

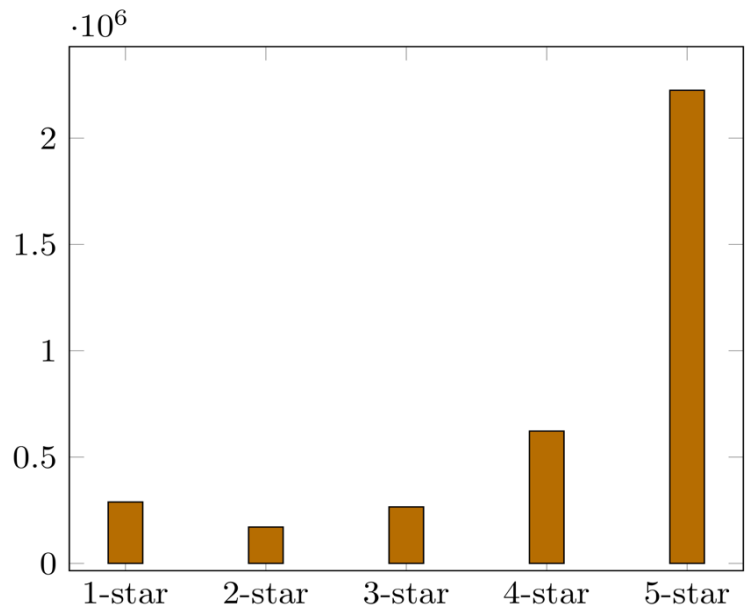

Figure 3 Distribution of reviews by rating

Online reviews contain a lot of noise like hyperlinks, html tags, informal words, etc. and many words don't have any significant impact on the sentiment of the review. Keeping such words in the review text will increase the dimensionality of the problem. To address this issue, we pre-processed each review before converting it to its corresponding vector. Our step-by-step approach included:

1. Removing hyperlinks.
2. Removing unwanted spaces between words.
3. Converting informal words such as 'I'll', 'I've' to its formal form 'I will', 'I have', etc.
4. Adding spaces between punctuation. For example, 'This is great!It works.' is converted to 'This is great ! It works .'. Punctuations are treated as separate tokens to try to improve the accuracy of the classifier.

## 2.2. MODELING OF REVIEWS USING PARAGRAPH VECTORS

Paragraph vectors (PV) [2] are highly inspired from the framework of word vectors [4]. Word vectors framework learns semantic relationships by predicting the next word from words in a given context. Similarly, PV framework learns vectors by predicting next word given many contexts that are sampled from a paragraph. There are two flavors of paragraph vector, a) Distributed memory model (PV-DM) b) Distributed bag of words model (PV-DBOW). The difference between the two models is that PV-DM captures word order while PV-DBOW ignores it.

Figure 4 illustrates the distributed memory model of the paragraph vector [2]. Every paragraph is mapped to a column of matrix D and similarly every word is mapped to a column of matrix W. Given a context sampled from a paragraph or document (for example $x_i$, $x_{i+1}$ and $x_{i+2}$), the model predicts the next word $x_{i+3}$ by concatenating the paragraph vector with vectors of words in that context. The model updates both the paragraph matrix and the word matrix while training to minimize error.

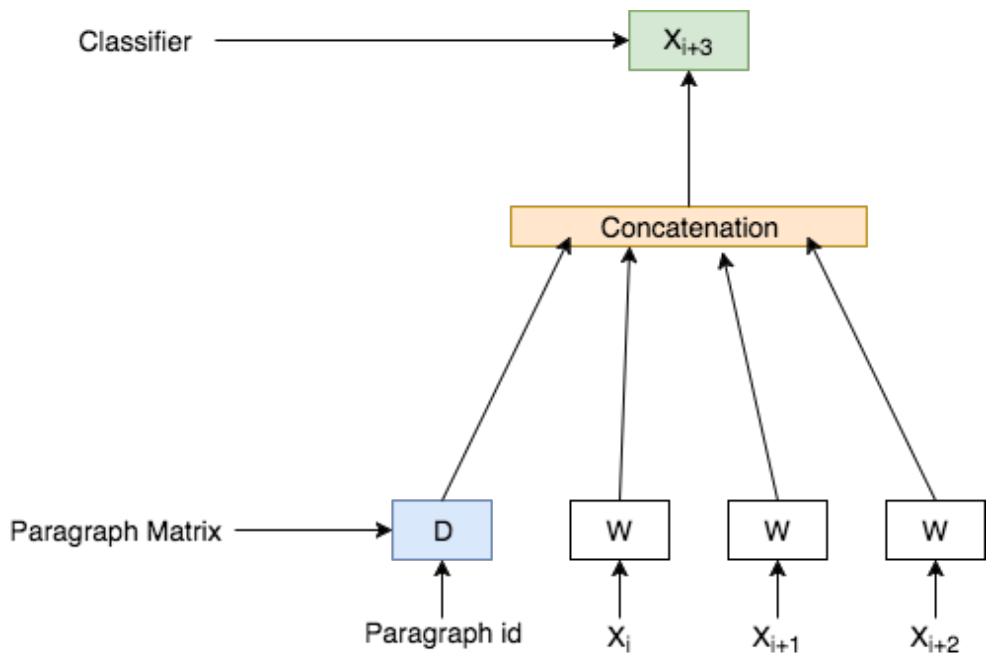

Figure 4 Distributed Memory Model of Paragraph Vector

Figure 5 illustrates the distributed bag of words model of paragraph vector [2]. This is different from PV-DM as it doesn't consider word ordering. PV-DBOW samples a random context from a paragraph and then a random word from that context, then based on that word it tries to predict the context. It doesn't use a word matrix and hence requires less data to store. According to the experiments carried out in [2], PV-DM is shown to be more efficient than PV-DBOW, therefore we have applied PV-DM to convert the reviews in our dataset to fixed length vectors.

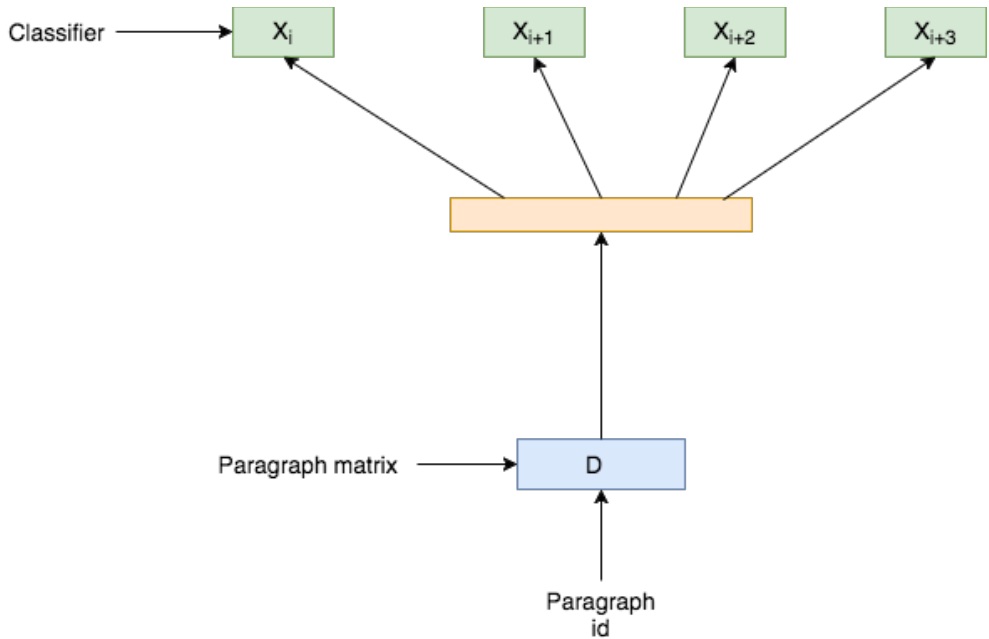

Figure 5 Distributed Bag of Words Model of Paragraph Vector

## 2.3. Learning Product Embeddings using Recurrent Neural Network with Gated Recurrent Unit

By using paragraph vectors, we converted 3.5 million product reviews to 300 dimensional fixed-length vectors. To compute product embeddings, we grouped the reviews by 'product id' and then ordered them by their 'review time'. Table 1 shows how a particular review is ordered for computing product embedding for that review. Review text is substituted by their respective paragraph vector. This temporally sorted vectors are the input sequence and their corresponding ratings are the targeted output and used to train a gated recurrent unit (GRU) to learn embeddings for that particular product.

Table 1 Review sequence for a product

| Product Id | Review | Review Time |
|---|---|---|
| 0060245867 | This is the first of several of this type and clearly the best of the group... | 2002-08-21 |
| 0060245867 | I loved the movies, however, I wasn't interested in merely "reading the movies" again. There was no need... | 2006-07-01 |
| 0060245867 | "If You Give a Mouse a Cookie" really has been the "It" book for some time. Parents love this story, ... | 2006-09-05 |

### 2.3.1. RECURRENT NEURAL NETWORKS

Traditional neural networks consider inputs to be independent of each other. But this approach is flawed for tasks such as predicting the next word in a sentence. Here, the next word depends upon the previous word or sequence of previous words. In such cases, RNNs are useful as they are great at capturing sequential information. RNNs have loops in their architecture that allow them to pass information they have collected from previous inputs while processing new input. Figure 6 shows a recurrent neural network unrolled through all time steps.

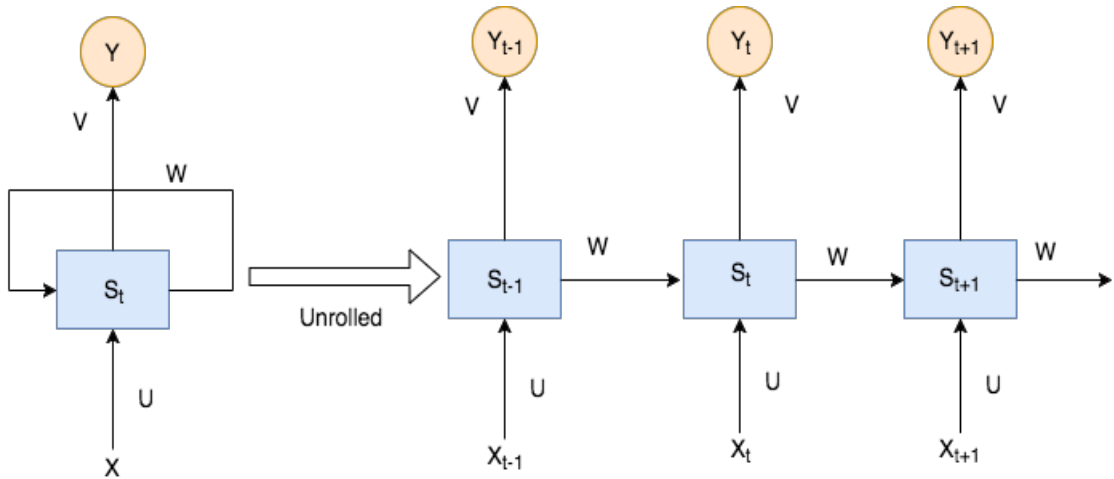

Figure 6 Unrolled RNN

In this architecture $X_t$ is input at time step t and $Y_t$ is output at the same time step. U, V and W are weight matrices that need to be learned and $S_t$ is the hidden state that is computed at every time step t. $S_t$ is also known as the memory of the network as it stores features about the input sequences based on its observation of previous inputs. For a given input sequence $x_1$, $x_2$, $x_3$, …, $x_T$, RNN computes sequence of hidden states and outputs using the Algorithm 1 shown below.

---
**Algorithm 1:** Computing hidden states and outputs of RNN

    **for** t=1 to T **do**
        $u_t \leftarrow U x_t + W s_{t-1} + b_h$
        $s_t \leftarrow f(u_t)$
        $o_t \leftarrow V s_t + b_o$
        $y_t \leftarrow g(o_t)$
    **end for**

---

$b_h$ and $b_o$ are bias vectors applied at hidden and output layers respectively. f and g are non-linear functions.

Recurrent neural networks are trained by gradient descent of error using an extension of back-propagation called back propagation through time (BPTT). Algorithm 2 shown below is used for computing gradients using BPTT.

---

**Algorithm 2:** Computing gradients using BPTT

---

$\quad$ **for** t=T to 1 **do**

$\qquad do_t \leftarrow g'(o_t) \cdot \frac{dL}{dy_t}$

$\qquad db_o \leftarrow db_o + do_t$

$\qquad dV \leftarrow dV + do_t s_t^T$

$\qquad ds_t \leftarrow ds_t + V^T do_t$

$\qquad dy_t \leftarrow f'(t_t) \cdot ds_t$

$\qquad dU \leftarrow dU + dy_t x_t^T$

$\qquad db_h \leftarrow db_h + dy_t$

$\qquad dW \leftarrow dW + dy_t \cdot ds_{t-1}^T$

$\qquad ds_{t-1} \leftarrow W^T \cdot dy_t$

$\qquad$ **return** $dV, dU, dW, db_o, db_h$

$\quad$ **end for**

---

### 2.3.2. RECURRENT NEURAL NETWORK WITH GATED FEEDBACK UNIT

When recurrent neural networks (RNN) are trained, the weight matrices U, V, and W are updated using backpropagation through time (BPTT). Update to each weight matrix is proportional to the gradient of the error with respect to that matrix and BPTT computes the gradients using the chain rule. When RNN is learning long-term context, depending on the activation functions, the gradient tends to get smaller (vanishing gradient problem) or bigger (exploding gradient problem) towards the earlier layers, which makes it difficult to train the network [15]. Gated Recurrent Units (GRU) were used to combat vanishing gradient problem through gating mechanism.

As shown in Table 1, for every product, the review vectors are sorted based on the review's posted time. These sorted review vectors along with their corresponding ratings were used as the training sequence for that particular product. These sequences were fed to a GRU to learn the 128-dimensional feature vector representation of product information. Training of GRU is performed using Algorithm 3 shown below.

---

**Algorithm 3:** Training GRU

---

$\quad$ **for** i=1 to Number of Epoch **do**

$\qquad$ **for** Sequence S in training sequences **do**

$\qquad\quad$ Train GRU with S

$\qquad\quad$ **if** New product or user sequence starts **then**

$\qquad\qquad$ Reset hidden states

$\qquad\quad$ **end if**

$\qquad$ **end for**

$\qquad$ Validate GRU with validation set

$\quad$ **end for**

---

As shown in Algorithm 3, the hidden states are reset after the start of a new product sequence since the objective is to capture information that exists between reviews belonging to a unique product sequence.

Output vector at the last time step of every product sequence is considered to be the embedding sequence for that particular product. We trained our GRU using 0.25 as dropout rate; Adam as stochastic optimization method; categorical cross entropy as loss function; time distributed dense layer; and 128 hidden units. During the training phase, product embedding sequence for each unique product was retrieved and stored in a file for efficient retrieval later on.

### 2.3.3. SENTIMENT CLASSIFICATION USING SUPPORT VECTOR MACHINES

Support vector machine (SVM) is a traditional machine learning algorithm to classify both linear and nonlinear data. Given a training data with binary outputs, support vector machine tries to find a hyper-plane as the decision surface such that the separation between positive and negative samples is maximized. If the data is non-linear, SVM transforms the data into a higher dimension and then solves the problem by finding a linear hyper-plane. The kernel used for such nonlinear data is called Gaussian Radial Basis function (RBF). SVM was used to classify each review to a label of "positive", "neutral", or "negative".

### 2.3.4. WEB SERVICE TO DETECT USER RATING MISMATCH

Figure 7 illustrates the web service we developed to prevent inconsistent review and rating pairs. It uses the support vector machine model trained with 3.5 million reviews and their product embeddings to predict the sentiment class for a given review.

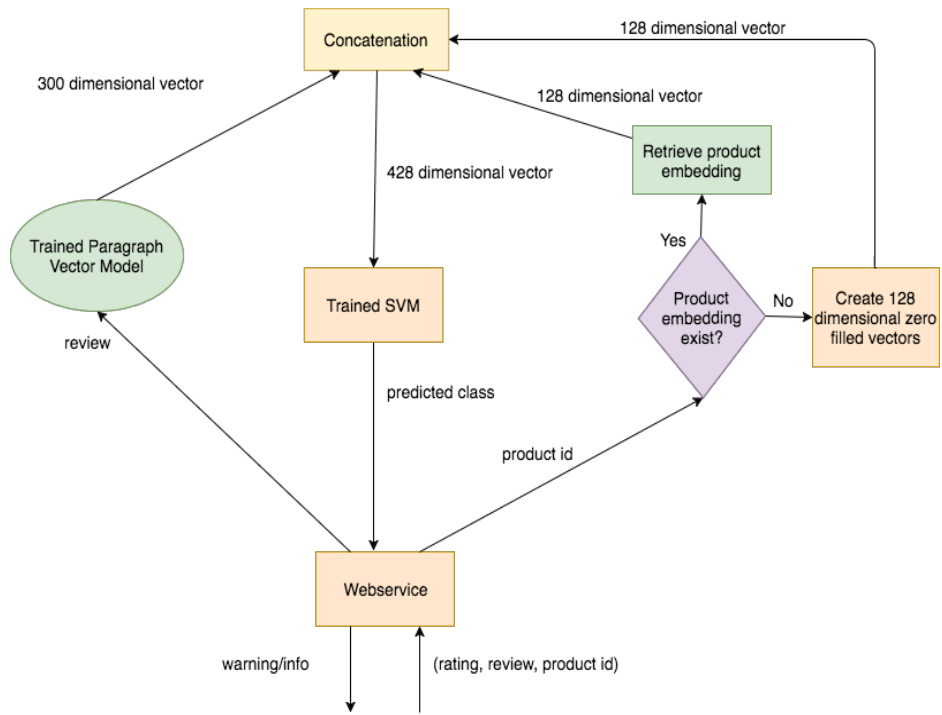

Figure 7 Web service architecture

When a user writes a review for a product and assigns a rating, if the predicted class and the rating class doesn't match, a feedback is provided to user so that if they wish, they can update their rating.

The web service takes three inputs: 'review text', 'assigned rating', and 'product id.' The 300-dimensional embedding for the 'review text' is predicted by the previously-trained paragraph vector model. As discussed earlier, product embeddings for all products are saved in a file. Embedding for the given 'product id' is retrieved by the web service from the file and concatenated with review embedding to form the final 428-dimensional vector. This vector is then used by our trained support vector machine model to predict a sentiment class. If there is mismatch between the predicted class and the sentiment class the assigned rating falls under, a feedback is given to the user.

Figure 8 shows a review with a positive sentiment but with a negative rating. There is a review-rating mismatch, therefore our system returns a message as shown in Figure 9. The user can then choose to either update the rating or keep it.

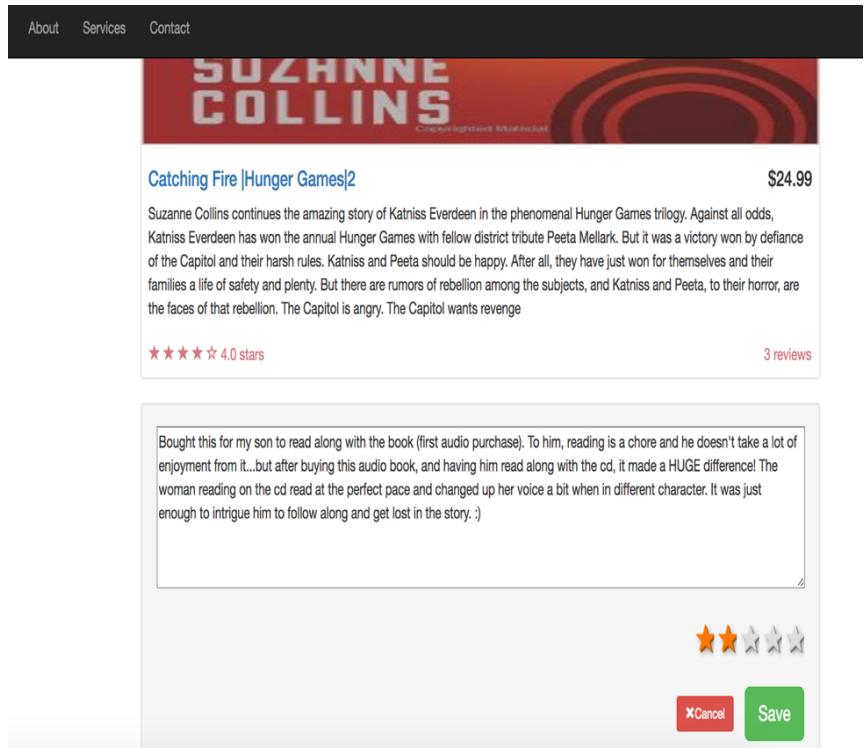

Figure 8 User interface for reviews

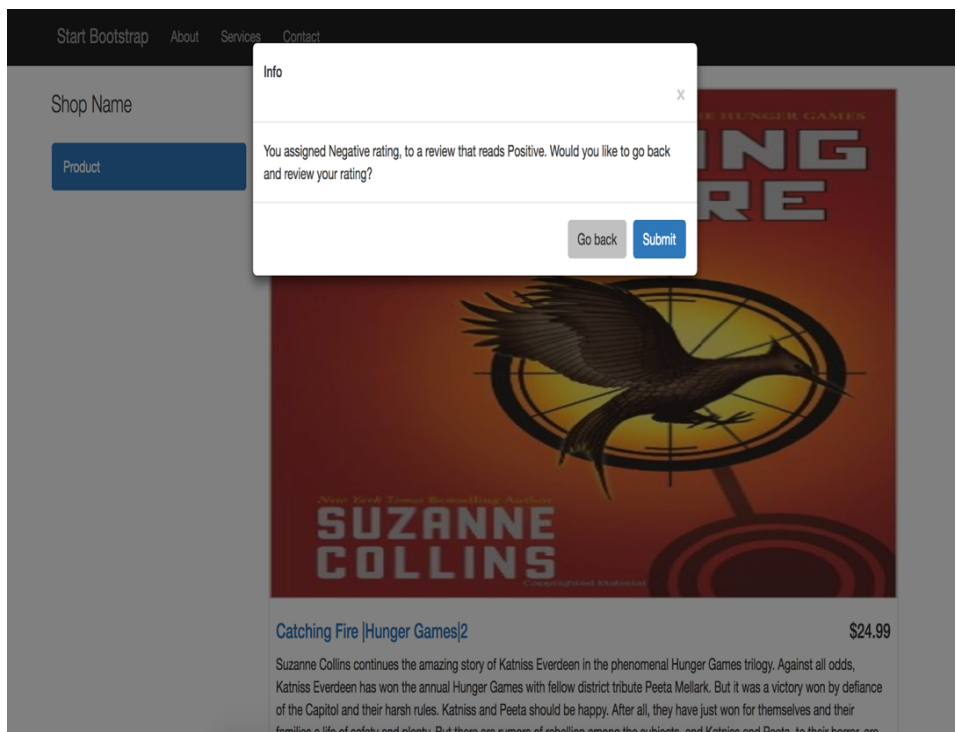

Figure 9 Feedback to the user

## 3. EXPERIMENTS AND RESULTS

### 3.1. SENTIMENT CLASSIFICATION USING REVIEW EMBEDDING

We first designed a paragraph vector only model using Genism framework. 3.5 million Amazon.com product reviews were converted to 300-dimensional fixed length vector. We trained this model for 15 epochs with learning rate of 0.025, window size of 10, and 35 worker threads. The learning rate was decreased by 0.002 after each epoch. This vector with its corresponding rating was used to train a support vector machine (SVM). 10-fold cross-validation method was used to evaluate the performance of this SVM. To evaluate the performance of our model, we computed precision, recall, and classification accuracy. Table 2 lists these metrics at each iteration as well as their average over 10 iterations. With only review embedding, the model had an average precision score of 0.5861, recall score of 0.4085, and classification accuracy of 81.29%.

Table 2 Precision, recall, and accuracy of the model with paragraph vectors only

| Iteration | Precision | Recall | Accuracy |
|-----------|-----------|--------|----------|
| 1 | 0.5890 | 0.4051 | 0.8125 |
| 2 | 0.5693 | 0.4072 | 0.8115 |
| 3 | 0.5817 | 0.4092 | 0.8128 |
| 4 | 0.6004 | 0.4091 | 0.8127 |
| 5 | 0.6028 | 0.4086 | 0.8135 |
| 6 | 0.5819 | 0.4088 | 0.8138 |
| 7 | 0.5731 | 0.4077 | 0.8139 |
| 8 | 0.5859 | 0.4097 | 0.8120 |
| 9 | 0.5841 | 0.4081 | 0.8123 |
| 10 | 0.5926 | 0.4115 | 0.8144 |
| **Average** | **0.5861** | **0.4085** | **0.8129** |

### 3.1. SENTIMENT CLASSIFICATION USING REVIEW EMBEDDING AND PRODUCT EMBEDDING

After generating the 300-dimensional embeddings for the product reviews, we grouped them by their 'product id' and sorted them by their 'review time.' Each group was an input to the gated recurrent unit. The embedding generated at the last time step of the sequence was treated as the embedding for that product. This product embedding was concatenated with the review embedding and used to train a support vector machine (SVM). 10-fold cross-validation was used to evaluate this model. We computed precision, recall and classification accuracy as we did with paragraph vector only approach. Table 3 lists these metrics at each iteration as well as their average over 10 iterations. With review embedding and product embedding, the model had an average precision score of 0.5945, recall score of 0.4252, and classification accuracy of 81.82%.

Table 3 Precision, recall, and accuracy of the model with paragraph vectors and GRU

| Iteration | Precision | Recall | Accuracy |
|-----------|-----------|--------|----------|
| 1 | 0.5987 | 0.4183 | 0.8168 |
| 2 | 0.5940 | 0.4314 | 0.8189 |
| 3 | 0.5915 | 0.4265 | 0.8185 |
| 4 | 0.5962 | 0.4252 | 0.8186 |
| 5 | 0.5844 | 0.4310 | 0.8179 |
| 6 | 0.5853 | 0.4231 | 0.8186 |
| 7 | 0.5951 | 0.4249 | 0.8180 |
| 8 | 0.6016 | 0.4301 | 0.8182 |
| 9 | 0.5925 | 0.4206 | 0.8179 |
| 10 | 0.6064 | 0.4211 | 0.8182 |
| **Average** | **0.5945** | **0.4252** | **0.8182** |

Classification using both review and product embeddings gave us an increase in classification accuracy from 81.29% to 81.82%, increase in precision from 0.5860 to 0.5945 and increase in recall from 0.408 to 0.4252. The performance of both the approach is compared in Figure 10.

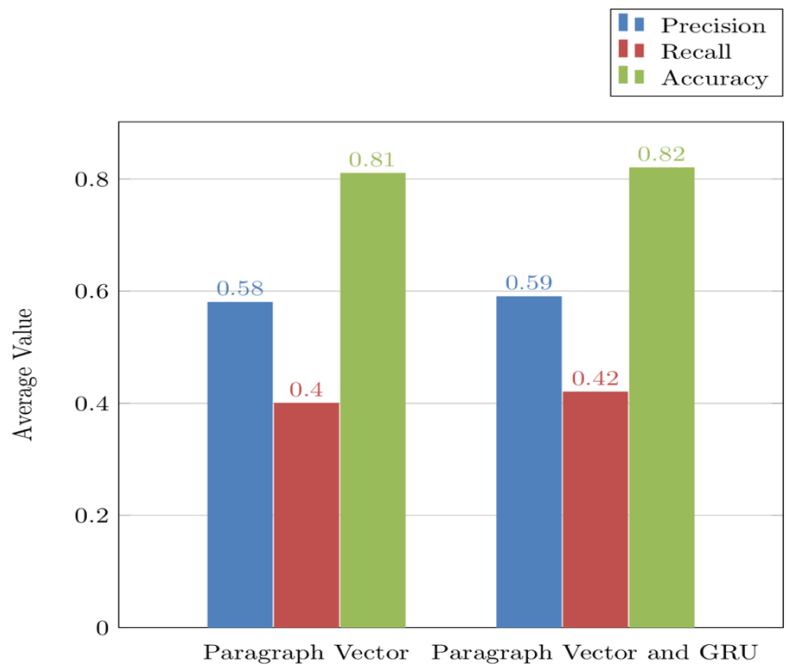

Figure 10 Comparison of metrics computed by review embedding only approach and both review embedding and product embedding approach

## 4. CONCLUSIONS

In this research, we propose a deep learning approach to solve a prevailing problem in e-commerce websites where the review submitted by the user doesn't match its rating. Reviews are a very important source of information for a potential customer before deciding to purchase a product [11]. We believe that such reviews with mismatched ratings can create negative user experience and hence the motivation for this research.

To construct our model, we first employed paragraph vectors to learn the syntactic and semantic relationship of a 'review text'. We further grouped and sorted review embedding to form a product sequence which is fed to a gated recurrent unit (GRU) to learn product embedding. The concatenation of review embedding generated from paragraph vectors and product embedding generated from GRU is used to train a support vector machine (SVM) for sentiment classification. With only review embedding our classifier performs at an accuracy of 81.29%. Inclusion of product embedding increases the accuracy to 81.82%. This shows that product information is a powerful feature that can be employed in sentiment analysis. We later use this classifier through a web service to predict rating of a review and compare it against given rating. This web service takes 'review text' and 'review rating' and provides a warning to the reviewer if there is an inconsistency between the given rating and review.

In summary, we showed how review text and product information together can be used to build a robust deep learning model that can be employed in sentiment analysis. We believe that a similar technique can be used to learn user information. Some users are more lenient and can give much higher ratings for the same review text compared to critical users [11]. We are motivated to exploit this information in future.